# Forecasting the Maxima of Solar Cycle 24 with Coronal Fe XIV Emission

**Richard C Altrock***

**Abstract** The onset of the "Rush to the Poles" of polar-crown prominences and their associated coronal emission is a harbinger of solar maximum. Altrock (*Solar Phys.* **216**, 343, 2003) showed that the "Rush" was well observed at 1.15 $R_o$ in the Fe XIV corona at the Sacramento Peak site of the National Solar Observatory prior to the maxima of Cycles 21 to 23. The data show that solar maximum in those cycles occurred when the center line of the Rush reached a critical latitude of $76° \pm 2°$. Furthermore, in the previous three cycles solar maximum occurred when the highest number of Fe XIV emission features per day (averaged over 365 days and both hemispheres) first reached latitudes $20° \pm 1.7°$. Applying the above conclusions to Cycle 24 is difficult due to the unusual nature of this cycle. Cycle 24 displays an intermittent Rush that is only well-defined in the northern hemisphere. In 2009 an initial slope of $4.6°$ year$^{-1}$ was found in the north, compared to an average of $9.4 \pm 1.7°$ year$^{-1}$ in the previous cycles. An early fit to the Rush would have reached $76°$ at 2014.6. However, in 2010 the slope increased to $7.5°$ year$^{-1}$ (an increase did not occur in the previous three cycles). Extending that rate to $76° \pm 2°$ indicates that the solar maximum in the northern hemisphere already occurred at $2011.6 \pm 0.3$. In the southern hemisphere the Rush to the Poles, if it exists, is very poorly defined. A linear fit to several maxima would reach $76°$ in the south at 2014.2. In 1999, persistent Fe XIV coronal emission known as the "extended solar cycle" appeared near $70°$ in the North and began migrating towards the equator at a rate 40 % slower than the previous two solar cycles. However, in 2009 and 2010 an acceleration occurred. Currently the greatest number of emission features is at $21°$ in the North and $24°$ in the South. This indicates that solar maximum is occurring now in the North but not yet in the South.

*****Air Force Research Laboratory**
**Space Weather Center of Excellence**
**PO Box 62**
**Sunspot, NM 88349 USA**
**altrock at nso.edu**

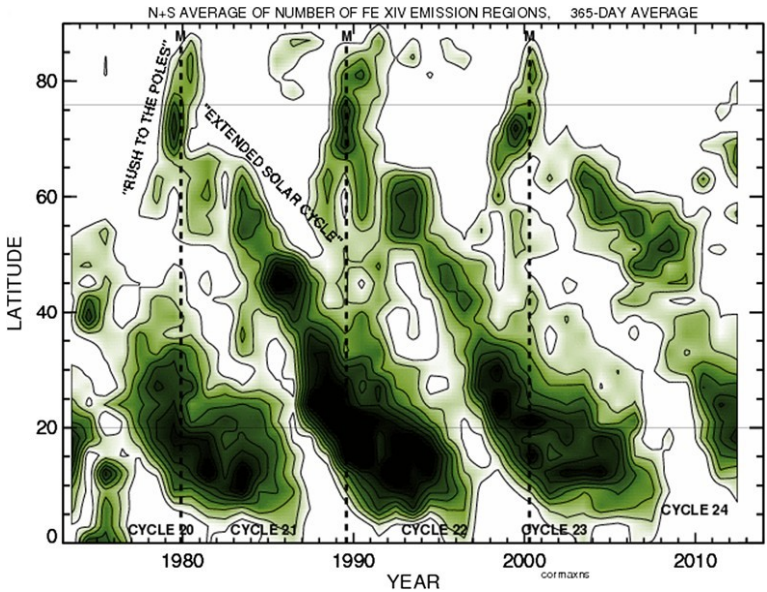

**Figure 1** Annual northern-plus-southern-hemisphere averages of the number of Fe XIV emission features from 1973 to 2012. The Rush to the Poles beginning near 1978 and the extended solar cycle beginning near 1980 (see text) are indicated. Vertical dashed lines indicate the time of solar maxima. Contours are drawn at 0.065, 0.085, 0.105, . . . emission features per day. Shading darkens with each change of contour. See text for description of other features.

## 1. Introduction

Altrock (2011) discussed in detail the methods by which observations of Fe XIV 530.3 nm obtained with the *Photoelectric Coronal Photometer* and 40 cm coronagraph at the John W. Evans Solar Facility of the National Solar Observatory at Sacramento Peak (Fisher 1973, 1974; Smartt, 1982) can be used to forecast solar maxima at varying times prior to the occurrence of the maxima. For further discussion of the observations see Altrock (1997) as well as earlier observations discussed in Altrock (2011).

Altrock (2011) found that a synoptic map of the location in latitude of local intensity maxima in the Fe XIV scans at 1.15 solar radii from the center of the disk ($R_o$) clearly showed the progress of the emission from high to low latitudes, known as the extended solar cycle, in Cycles 22 – 23 and the Rush to the Poles overlying polar-crown prominences preceding solar maxima. We will hereafter refer to these intensity maxima as emission features.

As discussed in Altrock (2011), high-latitude emission features are situated above the high-latitude neutral line of the large-scale photospheric magnetic field seen in Wilcox Solar Observatory synoptic maps. This is also the locus of polar-crown prominences. These features are therefore likely parts of streamers overlying the polar-crown prominences, although the low resolution of the observations does not allow a rigorous connection to be made. At lower latitudes the emission features may also overly active regions, prominences, other large-scale magnetic field boundaries, *etc*.

As in Altrock (2011), we average the number of points at each latitude over a given time interval. This process allows us to correct the data for missing days, which is an essential step for correctly interpreting the data. Figure 1 shows annual averages of the number of emission features, also averaged over the northern and southern hemispheres.

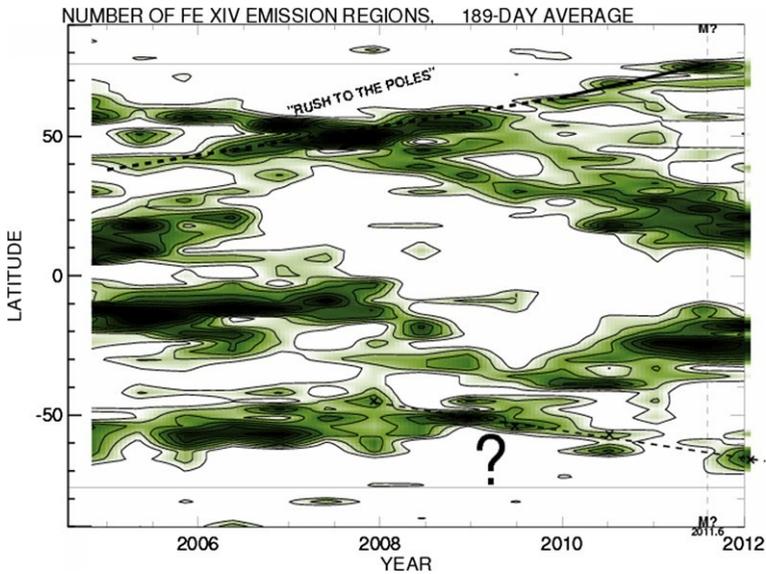

**Figure 2** Seven-rotation (approximately semiannual) averages of the number of Fe XIV emission features from 2005 to 2012 plotted separately for each hemisphere (−90º to +90º latitude). See text for further information.

## 2. Discussion

In Figure 1, we see that extended Solar Cycles 22 and 23 begin near $70°$ latitude and end near the equator about 18 years later. The initial rate of migration towards the equator for Cycle 24 was 40 % slower than the previous cycles (see Altrock, 2011). More recently, emission took a more rapid rate down to near $20°$ latitude, and there is a suggestion that a Rush may have developed.

Also in Figure 1, there are fine lines drawn at $76°$ and $20°$. As discussed in Altrock (2011), during the last three cycles, solar maximum occurred when a linear fit to the Rush reached $76° \pm 2°$ latitude. This may be seen in Figure 1. Altrock (2011) also found that, in the previous three cycles, solar maximum occurred when the greatest number of Fe XIV emission features, averaged over 365 days and both hemispheres (as in Figure 1), were at latitudes $20° \pm 1.7°$. Suggestions of this may also be seen in Figure 1.

In order to attempt to apply these conclusions to Cycle 24, let us examine a higher-resolution (if noisier) graphic.

### 2.1. The Rush to the Poles

Figure 2 shows the data plotted separately for each hemisphere from 2005 – 2012 (for earlier years, see Altrock (2011)). Fine lines mark $76°$ latitude. In 2005 there is an appearance of a weak Rush in the northern hemisphere, marked by a label, "Rush to the Poles", and a linear fit from 2005 – 2010 (dashed line). This early fit to the Rush resulted in more of a stroll than a Rush, which would have reached $76°$ at 2014.6. Between 2009 and 2010 an acceleration occurred (solid line). The new, higher rate of the Rush in the northern hemisphere indicates it reached $76° \pm 2°$ latitude at 2011.6 ± 0.3. In other words, in the northern hemisphere, solar maximum has already occurred at about 2011.6. This is indicated by a fine vertical dashed line.

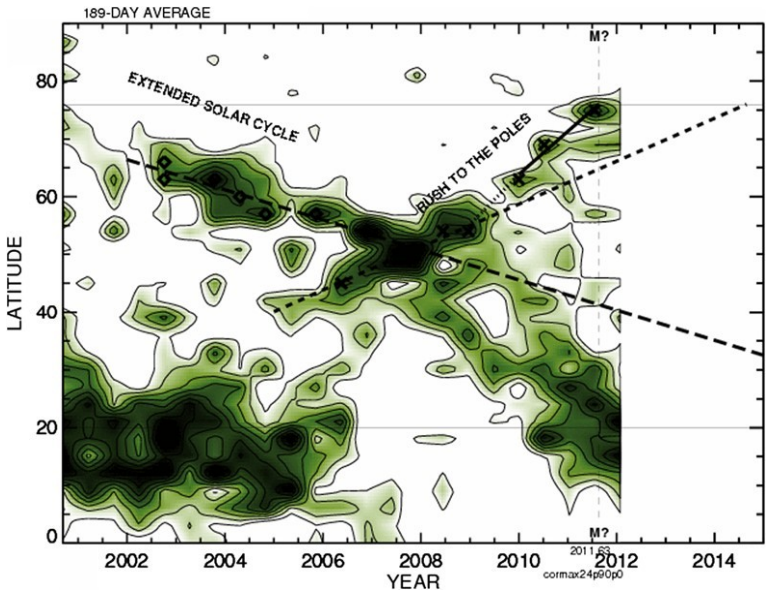

**Figure 3** Seven-rotation averages of Cycle 24 northern-hemisphere emission features.

In the southern hemisphere, the Rush to the Poles, if it exists, is very poorly defined. I show a linear fit to several maxima with a dashed line underlain by a large interrogative. This fit would reach 76° at 2014.2, which gives a best guess to solar maximum in the southern hemisphere.

2.2. Low-Latitude Emission

Figure 3 shows the northern-hemisphere emission from 2000.5 to 2012. The northern-hemisphere fits to the Rush discussed in Section 2.1 are shown, as well as the initial fit to the extended solar cycle, which is extrapolated to 2015. This initial fit (long-dashed line) would not reach 20° until 2019.8.

However, the initial rate of migration towards the equator of emission features was suddenly interrupted around 2009. The higher-latitude emission is ending, and a new lower-latitude band has developed. This low-latitude band may be used to infer when solar maximum will occur. Currently the greatest number of emission features is at 21° in the North and 24° in the South. This indicates that solar maximum is occurring now in the North but not yet in the South.

Note, from Sections 2.1 and 2.2, that something remarkable happened on the Sun around 2009, resulting in a dramatic increase in the latitude-time rates in the northern hemisphere of the extended solar cycle and the Rush to the Poles.

2.3. Results from Other Authors and Other Data Sets

Gopalswamy *et al.* (2012) come to similar conclusions from a study of the microwave butterfly diagram and prominence eruption latitudes and show that the northern high-latitude magnetic field changes sign in 2012, which is an indicator that solar maximum has occurred in the North. Rušin, Minarovjech, and Saniga (2009), using similar Fe XIV data but a different technique, predicted two solar maxima: one in the time frame 2010 – 2011 and one in 2012.

Other authors have discussed North-South asymmetries. Gopalswamy *et al.* (2003) showed that in Cycle 23 the magnetic polarity of the north pole reversed before that of the south pole. This is consistent with the observations that the northern polar-crown prominences disappeared in October of 2000, while those in the South disappeared in early 2002, and northern high-latitude activity disappeared in November 2000, while that in the South disappeared in May 2002.

Svalgaard and Kamide (2012) note that in Cycle 19 the south pole reversed polarity first followed by the north pole more than a year later. Since then polar field reversals have happened first in the North. They show that North-South asymmetries are normal in the sunspot number (each hemisphere may have two or more maxima), and the Rush to the Poles for Cycle 20 happened in the North well before the South.

Gopalswamy (2012) studied solar energetic particle (SEP) events for Cycle 24 and found that all but one of the 15 large SEP events during the first 4.5 years occurred in the northern hemisphere, and SEP events did not appear in the southern hemisphere until June 2012. Cycle 24 therefore resembles the rise and maximum phases of Cycle 21, when most of the events occurred in the northern hemisphere until after the maximum phase.

Altrock (2011) discusses the use of simulated coronal emission *vs.* observed coronal emission. He concludes that there is no reason to use simulated emission in studies of the properties of coronal emission. Robbrecht *et al.* (2010) used potential-field source-surface simulated emission to conclude that the extended solar cycle does not exist. However, other authors (*cf*. Sandman and Aschwanden, 2011) doubt the validity of this simple model.

The usual measure of solar maximum, the global smoothed sunspot numbers (ftp://ftp.ngdc.noaa.gov/STP/SOLAR_DATA/SUNSPOT_NUMBERS/INTERNATIONAL/smoothed/SMOOTHED), show an inflection point in late 2011, which could represent solar maximum in the northern hemisphere.

Figure 4 shows North and South sunspot areas as compiled at NASA Marshall SFC by David Hathaway (http://solarscience.msfc.nasa.gov/greenwch.shtml) from 2007 to 2012. These monthly values have been smoothed in the same manner as the global smoothed sunspot numbers but are extended up to the current time by use of IDL edge truncation. The last value unaffected by edge truncation is January 2012. However, it seems all but certain that solar maximum has occurred in the North but not yet in the South, albeit in the sunspot *area*, not in the sunspot *number*. The Solar Influences Data Analysis Center (SIDC) (http://www.sidc.be/) has northern and southern sunspot numbers. Their data show a maximum in the northern hemisphere near 2012.0 but no maximum yet in the South.

## 3. Conclusions

The location of Fe XIV emission features in time-latitude space displays an 18-year progression from near $70°$ to the equator, which has been referred to as the extended solar cycle. Cycle 24 emission features began migrating towards the equator similarly to previous cycles, although at a 40 % slower rate. In addition, in approximately 2005 the northern-hemisphere Rush to the Poles began at a 50 % slower rate than in recent cycles. It accelerated beginning in 2010.

Analysis of the northern-hemisphere Rush to the Poles indicates that the northern-hemisphere maximum already occurred in 2011.6 ± 0.3. This conclusion is different from that in Altrock (2011), because the acceleration beginning in 2010 had not yet been recognized.

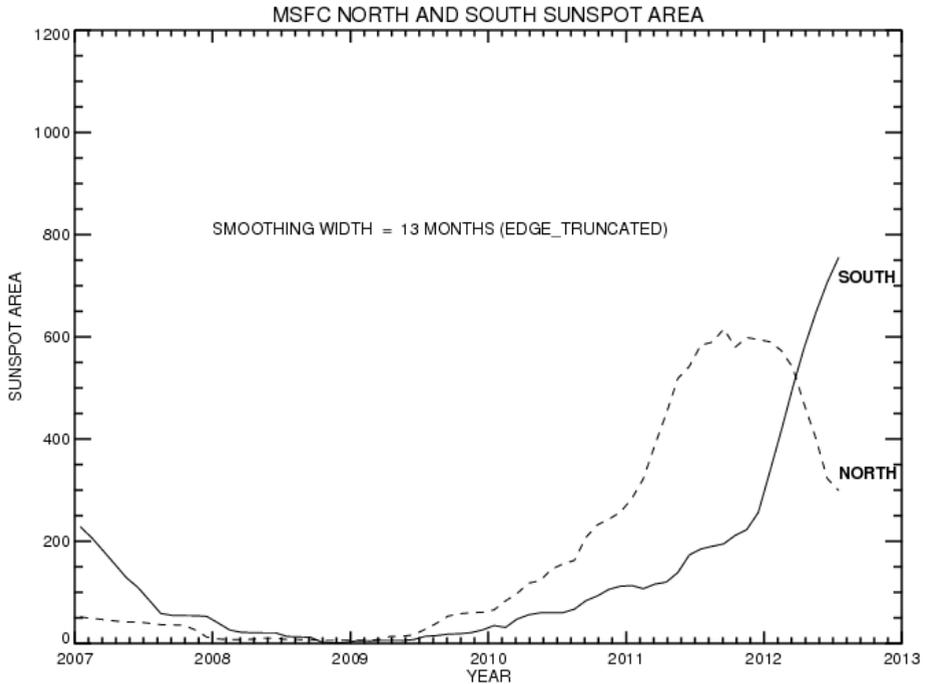

**Figure 4** Smoothed NASA Marshall SFC northern and southern sunspot areas.

A weak Rush may be occurring in the southern hemisphere, and an analysis of its properties yields an estimate for the southern-hemisphere maximum of 2014.2.

Low-latitude emission features are migrating towards the equator in both hemispheres. In previous cycles, solar maximum occurred when the greatest concentration of 1.15 $R_o$ Fe XIV emission features reached $20°  ± 1.7°$ latitude. Currently the greatest number of emission features is at $21°$ in the North and $24°$ in the South. This indicates that solar maximum is occurring now in the North but not yet in the South.

A recent study of microwave brightness and prominence eruptions by Gopalswamy *et al.* (2012) comes to similar conclusions. In addition, there is strong evidence that the northern-hemisphere sunspot *area* reached a maximum around the end of 2011. The southern-hemisphere area appears to still be increasing. An inflection point occurred in the global smoothed sunspot *number* late in 2011, which could be evidence that maximum occurred in one hemisphere.

This analysis does not indicate the strength of the maximum.

**Acknowledgments**   The observations used herein are the result of a cooperative program of the Air Force Research Laboratory and the National Solar Observatory. I am grateful for the assistance of NSO personnel, especially John Cornett, Timothy Henry, Lou Gilliam, and Wayne Jones, for observing and data-reduction and -analysis services and maintenance of the Evans Solar Facility and its instrumentation and to Raymond N. Smartt, who completely redesigned the Sacramento Peak *Photoelectric Coronal Photometer* filters in 1982, making it the excellent instrument it is today. I wish to also thank the referee for pointing out several deficiencies in the paper, which I have endeavored to correct.